\documentclass[useAMS, usenatbib]{mn2e}
\usepackage{graphicx}
\usepackage{color}
\usepackage{tabularx}
\usepackage{rotating}
\usepackage{url}
\usepackage{longtable}
\usepackage{supertabular}
\usepackage{multirow}
\usepackage{scrextend}

\begin{document}
\title[The influence of molecular outflows on GB clouds]
	{The JCMT Gould Belt Survey:  Understanding the influence of molecular outflows on Gould Belt clouds}
	
\author[E. Drabek-Maunder et al.]{E. Drabek-Maunder$^{1}$\thanks{E-mail: e.drabek-maunder@imperial.ac.uk}, J. Hatchell$^{2}$, J. V. Buckle$^{3,4}$, J. Di Francesco$^{5,6}$, J. Richer$^{3,4}$\\
$^{1}$Imperial College London, Blackett Laboratory, Prince Consort Rd, London SW7 2BB\\
$^{2}$University of Exeter, School of Physics, Stocker Road, Exeter EX4 4QL\\
$^{3}$Astrophysics Group, Cavendish Laboratory, J J Thomson Avenue, Cambridge, CB3 0HE\\
$^{4}$Kavli Institute for Cosmology, Institute of Astronomy, University of Cambridge, Madingley Road, Cambridge, CB3 0HA, UK\\
$^{5}$NRC Herzberg Astronomy and Astrophysics, 5071 West Saanich Rd, Victoria, BC, V9E 2E7, Canada\\
$^{6}$Department of Physics and Astronomy, University of Victoria, Victoria, BC, V8P 1A1, Canada\\
}

\maketitle

\begin{abstract}

Using JCMT Gould Belt Survey data from CO $J=3\rightarrow2$ isotopologues, we present a meta-analysis of the outflows and  energetics of star-forming regions in several Gould Belt clouds.  The majority of the regions are strongly gravitationally bound. There is evidence that molecular outflows transport large quantities of momentum and energy.  Outflow energies are at least 20~per~cent of the total turbulent kinetic energies in all of the regions studied and greater than the turbulent energy in half of the regions.  However, we find no evidence that outflows increase levels of turbulence, and there is no correlation between the outflow and turbulent energies.  Even though outflows in some regions contribute significantly to maintaining turbulence levels against dissipation, this relies on outflows efficiently coupling to bulk motions.  Other mechanisms (e.g. supernovae) must be the main drivers of turbulence in most if not all of these regions.


%

\end{abstract}



\section{Introduction}


Molecular outflows are expected to have various roles in the star formation process. Individual outflows not only provide a record of protostellar mass-loss, but may also carry away excess angular momentum so that mass can accrete onto the central protostars \citep{2002ApJ...576..222B}.  Theoretical work has focused on understanding how outflows influence their environments (e.g., \citealt{2007ApJ...662..395N, 2007ApJ...659.1394M, 2006ApJ...640L.187L}).  Even though high-mass young stellar objects (YSOs) have larger and more powerful outflows, populations of low-mass YSOs may be equally disruptive by interacting with a sizeable portion of their environment.  Energy input by outflows can be comparable to or larger than cloud turbulent and gravitational energies (e.g., \citealt{2010MNRAS.401..204B, 2010MNRAS.408.1516C, 2010MNRAS.409.1412G}).  If outflows are well coupled to the cloud, they may act as a considerable or dominant source of turbulence and provide global support against gravitational collapse \citep{2004RvMP...76..125M}.


In this Letter, we present a meta-analysis using James Clerk Maxwell Telescope (JCMT) Gould Belt Legacy Survey (GBS; \citealt{2007PASP..119..855W}) data to give a first look at the momenta and energetics of high-velocity outflows and compare their influence on star-forming regions.  This brings together previous work that used Heterodyne Array Receiver Programme (HARP; \citealt{2009MNRAS.399.1026B}) to observe CO~$J=3\rightarrow2$ transitions and to analyse the mass and energetics of ambient and outflowing gas in Perseus regions (\citealt{2010MNRAS.408.1516C}; C10), Serpens Main (\citealt{2010MNRAS.409.1412G}; G10) and Ophiuchus L1688 (\citealt{2015MNRAS.447.1996W}; W15).

\section{Method}
\label{method}

We base our meta-analysis on GBS observations of the $J=3\rightarrow2$ transitions of $^{12}$CO (345.7960~GHz), $^{13}$CO (330.5880~GHz) and  C$^{18}$O (329.3306~GHz).  Details of the data reduction process and calculations of the ambient gas and outflow properties are given in C10 (Perseus), G10 (Serpens Main) and W15 (Oph L1688).  The GBS utilised the same emission lines in each region, which helps ensure the ambient gas and outflow properties are being traced consistently.  However, the methods used to determine the mass and energetics differed slightly between the studies as we now describe.  The regional mass and energetics were calculated from C$^{18}$O line emission (or $^{13}$CO for L1455).  In L1688 (W15), the mass and energetics were corrected for high C$^{18}$O optical depths.  C$^{18}$O was assumed to be optically thin for Perseus regions (NGC~1333, IC~348, L1448 and L1455; C10) and Serpens Main (G10), so mass, momentum and energy are lower limits.  Additionally, radii for both L1688 and Serpens Main were calculated from an effective cloud radius (determined from the total areas detected in C$^{18}$O), whereas radii for Perseus regions were taken to be geometric averages of the major and minor axes.  Lastly, line widths for both L1688 and Serpens Main regions were calculated using the average C$^{18}$O spectra, neglecting thermal line widths, but non-thermal line widths for Perseus were calculated from individual spectra measured at each position and averaged across the maps.  This latter method led to lower line width estimates (due to the size-linewidth relation; \citealt{1981MNRAS.194..809L}), leading to lower turbulent energies.  

In GBS studies, outflow properties were determined from $^{12}$CO emission assuming a 50~K gas temperature.  Oph L1688 and Serpens Main outflow properties were evaluated from high-velocity emission integrated over the mapped regions (red- and blue-shifted from the line centre).  This method can potentially include high-velocity emission driven by other sources (e.g., winds from nearby OB associations).  Conversely, an individual outflow-by-outflow analysis was used to determine the Perseus outflow properties, which could potentially miss emission not in the immediate vicinity of an outflow lobe.  Additionally, outflow properties in L1688 and the Perseus regions were corrected for optically thick $^{12}$CO emission.  Outflow mass and energetics in Serpens Main, however, should be considered a lower limit since the $^{12}$CO emission was assumed to be optically thin.  Lastly, outflow properties were also corrected for random inclination.

Past work (e.g., \citealt{2010ApJ...710L.142F, 2014ApJ...783..115N}) indicates outflow feedback is momentum rather than energy-driven because clumps are expected to have efficient energy loss.  Here, we examine both the momentum and energy transport from outflows to turbulence to better understand how outflows influence their environment.  The rates at which outflows inject momentum and energy into the ambient gas are $dP_\mathrm{out}/dt = P_\mathrm{out}/T_\mathrm{I}$ and $dE_\mathrm{out}/dt = E_\mathrm{out}/T_I$, where $P_\mathrm{out}$ is the outflow momentum, $E_\mathrm{out}$ is the outflow kinetic energy and $T_\mathrm{I}$ is the typical outflow lifetime.  Past work (e.g., \citealt{1991MNRAS.252..442P}) found that dynamical timescales can underestimate the outflow duration by an order of magnitude.  Therefore, we use the average lifetime of a Class~I protostar ($\sim0.5$~Myr; \citealt{Evans09}) as protostars are observed to produce outflows from the start of the Class~0 until the end of the Class~I stage.  In turn, the dissipation rates of the momentum and energy supersonic turbulence are calculated as $dP_\mathrm{turb}/{dt} = (0.21) {M \mathrm{\sigma_{3D}}}/{(\lambda_\mathrm{d}/\mathrm{\sigma_{3D}})}$ and ${dE_\mathrm{turb}}/{dt} =  {(0.42) M\mathrm{\sigma_{3D}^2}}/{(\lambda_d/\mathrm{\sigma_{3D}})}$, respectively, where $\sigma_\mathrm{3D}$ is the three-dimensional velocity dispersion of the C$^{18}$O ambient gas, $\lambda_\mathrm{d}$ is the driving length scale and $M$ is the mass of the cloud (see \citealt{1999ApJ...524..169M, 2014ApJ...783..115N}).

\section{Results}
\label{results}

\begin{table*}
\fontsize{8}{8}\selectfont
\begin{minipage}{7in}
\centering
\caption{Data for Perseus (C10), Serpens Main (G10) and Oph L1688 (W15).   The energetics of Perseus regions have been corrected for consistency with other regions.  Distance uncertainties are based on aforementioned work and references therein (including \citealt{2010ApJ...715.1170A} for Perseus) and are used for estimating mass and energetics uncertainties.  We derive two values for the Serpens Main mass and energetics based on differing distance estimates: (1) 230~pc (G10) and (2) 429~pc \citep{2011RMxAC..40..231D}.   }
\renewcommand{\arraystretch}{1.2}
\begin{tabular}{c c c c c c c c}
\hline
Cloud & Distance & Radius & $\sigma_{\mathrm{3D}}$ & Mass & $E_\mathrm{grav}$ & $E_\mathrm{turb}$  & $E_{out}$   \\
& (pc) & (pc) & (km~s$^{-1}$) & (M$_\odot$) & ($\mathrm{M_\odot}$~km$^2$~s$^{-2}$) &  ($\mathrm{M_\odot}$~km$^2$~s$^{-2}$) &  ($\mathrm{M_\odot}$~km$^2$~s$^{-2}$)  \\
\hline
NGC~1333 & 250$\left(^{+70}_{-50}\right)$ & 0.94 & 0.76 & 439$\left(^{+280}_{-158}\right)$ & 1761$\left(^{+1932}_{-859}\right)$ & 128$\left(^{+82}_{-46}\right)$ & 246$\left(^{+157}_{-89}\right)$\\
IC~348 & 250$\left(^{+70}_{-50}\right)$ & 0.53 & 0.45 & 196$\left(^{+125}_{-71}\right)$ & 604$\left(^{+663}_{-295}\right)$ & 20$\left(^{+13}_{-7}\right)$ & 5$\left(^{+3}_{-2}\right)$ \\ 
L1448 & 250$\left(^{+70}_{-50}\right)$ & 0.24 & 0.61 & 59$\left(^{+38}_{-21}\right)$ & 126$\left(^{+138}_{-61}\right)$ & 11$\left(^{+7}_{-4}\right)$ & 272$\left(^{+174}_{-98}\right)$  \\
L1455 & 250$\left(^{+70}_{-50}\right)$ & 0.11 & 0.45 & 19$\left(^{+12}_{-7}\right)$ & 28$\left(^{+31}_{-14}\right)$ & 3$\left(^{+2}_{-1}\right)$ & 8$\left(^{+5}_{-3}\right)$   \\
Serp Main (1) & 230$\left(^{+20}_{-20}\right)$ & 0.35 & 1.47 & 203$\left(^{+37}_{-34}\right)$ & 246$\left(^{+77}_{-54}\right)$ & 221$\left(^{+40}_{-37}\right)$ & 151$\left(^{+27}_{-25}\right)$ \\
Serp Main (2) & 429$\left(^{+2}_{-2}\right)$ & 0.35 & 1.47 & 706$\left(^{+7}_{-7}\right)$ & 1596$\left(^{+22}_{-22}\right)$ & 769$\left(^{+7}_{-7}\right)$ & 525$\left(^{+5}_{-5}\right)$ \\
Oph L1688 & 120$\left(^{+40}_{-4}\right)$ & 0.50 & 1.11 & 515$\left(^{+400}_{-35}\right)$ & 2264$\left(^{+3103}_{-219}\right)$ & 317$\left(^{+246}_{-22}\right)$ & 65$\left(^{+50}_{-4}\right)$ \\
\hline
\end{tabular}
\label{table1}
\end{minipage}
\end{table*}

\begin{table*}
\fontsize{8}{8}\selectfont
\centering
\caption{Rates that momentum and energy are injected into the cloud from outflows and dissipated through turbulence.}
\renewcommand{\arraystretch}{1.2}
\begin{tabular}{c c c c c}
\hline
Cloud & $dP_\mathrm{out}/dt$ & $dP_\mathrm{turb}/dt$ & $dE_\mathrm{out}/dt$ & $dE_\mathrm{turb}/dt$ \\
& ($\mathrm{M_\odot}$~km~s$^{-1}$/yr) & ($\mathrm{M_\odot}$~km~s$^{-1}$/yr) & $(\mathrm{M_\odot}$~km$^2$~s$^{-2}$/yr) & ($\mathrm{M_\odot}$~km$^2$~s$^{-2}$/yr) \\
\hline
NGC~1333 & $3.9\left(^{+2.5}_{-1.4}\right)\times10^{-5}$ & $0.5\left(^{+0.1}_{-0.1}\right)$--$2.7\left(^{+0.8}_{-0.5}\right)\times10^{-4}$ & 4.9$\left(^{+3.1}_{-1.8}\right)\times10^{-4}$ & 0.9$\left(^{+0.3}_{-0.2}\right)$--4.1$\left(^{+1.1}_{-0.8}\right)\times10^{-4}$ \\
IC~348 & $1.0\left(^{+0.6}_{-0.4}\right)\times10^{-5}$ & 1.6$\left(^{+0.4}_{-0.3}\right)$--4.3$\left(^{+1.2}_{-0.9}\right)\times10^{-5}$ & 1.0$\left(^{+0.6}_{-0.5}\right)\times10^{-5}$ & 1.4$\left(^{+0.4}_{-0.3}\right)$--3.8$\left(^{+1.1}_{-0.8}\right)\times10^{-5}$ \\ 
L1448 & $2.7\left(^{+1.7}_{-1.0}\right)\times10^{-5}$ & 2.0$\left(^{+0.6}_{-0.4}\right)$--2.4$\left(^{+0.7}_{-0.5}\right)\times10^{-5}$ & 5.4$\left(^{+3.5}_{-1.9}\right)\times10^{-4}$ & 2.3$\left(^{+0.6}_{-0.5}\right)$--2.7$\left(^{+0.8}_{-0.5}\right)\times10^{-5}$ \\
L1455 & $1.8\left(^{+1.1}_{-0.6}\right)\times10^{-5}$ & 5.9$\left(^{+1.7}_{-1.2}\right)\times10^{-6}$ & 1.5$\left(^{+1.0}_{-0.5}\right)\times10^{-5}$ & 1.1$\left(^{+0.3}_{-0.2}\right)\times10^{-5}$  \\
Serp Main (1) & $5.0\left(^{+0.9}_{-0.9}\right)\times10^{-5}$ & 2.7$\left(^{+0.2}_{-0.2}\right)$--4.7$\left(^{+0.4}_{-0.4}\right)\times10^{-4}$ & 3.1$\left(^{+0.6}_{-0.5}\right)\times10^{-4}$ & 0.8$\left(^{+0.1}_{-0.1}\right)$--1.4$\left(^{+0.1}_{-0.1}\right)\times10^{-3}$  \\
Serp Main (2) & $17.4\left(^{+0.2}_{-0.2}\right)\times10^{-5}$ & 5.0$\left(^{+0.1}_{-0.1}\right)$--8.8$\left(^{+0.1}_{-0.1}\right)\times10^{-4}$ & 10.8$\left(^{+0.1}_{-0.1}\right)\times10^{-4}$ & 1.5$\left(^{+0.1}_{-0.1}\right)$--2.6$\left(^{+0.1}_{-0.1}\right)\times10^{-3}$  \\
Oph L1688 & $5.6\left(^{+4.4}_{-0.4}\right)\times10^{-5}$ & 2.7$\left(^{+0.9}_{-0.1}\right)$--6.8$\left(^{+2.3}_{-0.2}\right)\times10^{-4}$ & 1.2$\left(^{+0.8}_{-0.1}\right)\times10^{-4}$ & 0.6$\left(^{+0.2}_{-0.02}\right)$--1.5$\left(^{+0.5}_{-0.1}\right)\times10^{-3}$ \\
\hline
\end{tabular}
\label{table2}
\end{table*}


\begin{figure}
\centering
\includegraphics[width=2.7in]{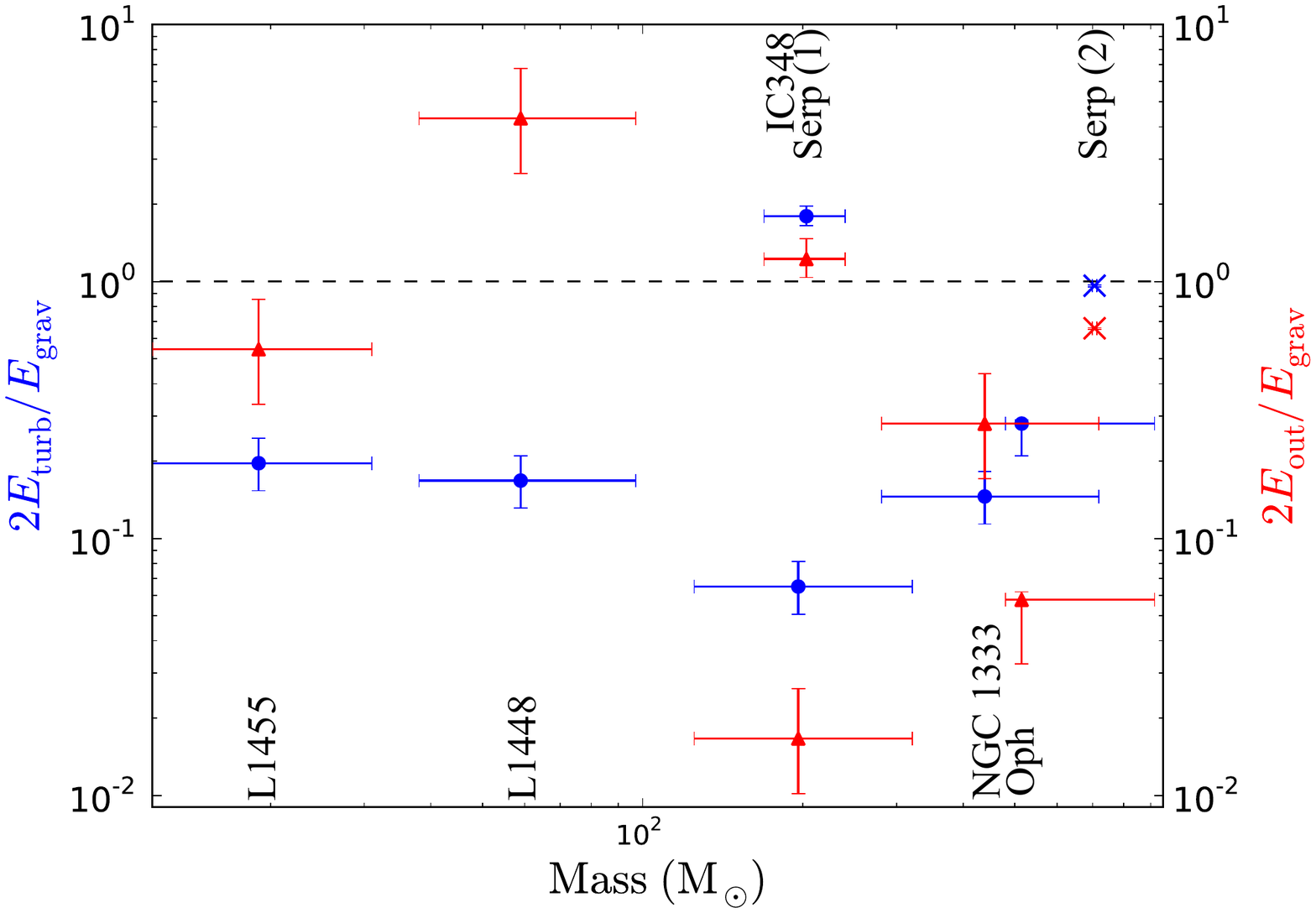}
\caption{Comparison between virial parameters, calculated from $E_\mathrm{grav}$ and $E_\mathrm{turb}$ (blue `$\bullet$') and $E_\mathrm{out}$ (red `$+$'), and mass.  Serp (2), denoted as `$\times$', is calculated using a 429~pc distance.}
\label{fig:mass_virial}
\end{figure}

Ambient gas and outflow calculations are presented in Tables~\ref{table1} and \ref{table2}.   The regions are diverse, spanning two orders of magnitude in outflow masses and energies.  Uncertainties on the mass and energetics have been calculated from distance uncertainties.  The largest systematic errors likely result from assuming constant gas temperatures and abundances for CO~$J=3\rightarrow2$ isotopologues (see C10, G10 and W15 for details).  Two distance estimates have been used for Serpens Main: (1) 230~pc (G10) and (2) 429~pc as from parallax measurements \citep{2011RMxAC..40..231D}.     


\subsection{Virial Parameter}

First, we investigate the regional stability using the virial parameter.  In Figure~\ref{fig:mass_virial}, we plot the virial parameter from each region's turbulent kinetic and gravitational binding energies ($2E_\mathrm{turb}/|E_\mathrm{grav}|$) and the ratio of the outflowing gas kinetic energy to the regional gravitational binding energy ($2 E_\mathrm{out}/|E_\mathrm{grav}|$).  The outflow kinetic energy cannot be used solely to assess if a region is in virial equilibrium because outflows do not necessarily contribute directly to the gravitational support.  This comparison can indicate if outflows are strong enough to overcome the local binding energy.   


All of the regions are bound with virial parameters $2E_\mathrm{turb}/|E_\mathrm{grav}|\leq0.3$ except for the marginally bound Serpens Main region ($2\mathrm{E_{turb}/|E_{grav}|}\sim1-2$; see \citealt{1992ApJ...395..140B}).  If there are no additional supporting forces (e.g. magnetic fields), then the low turbulent kinetic energies could indicate the regions are undergoing global collapse. 

Regions with relatively strong outflow kinetic to gravitational energies include Serpens Main, NGC~1333, L1455 and L1448.  In particular, L1448 has an outflow energy that surpasses its binding energy ($2 E_\mathrm{out}/|E_\mathrm{grav}|\sim4$), and C10 suggest there is potential for outflows to disperse the ambient gas if they are significantly coupled to the gas.  In Serpens Main, G10 suggest the high outflow and turbulent energies could indicate that outflows are the main driver of turbulence, causing the region to be near virial equilibrium.  Similarly, both NGC~1333 and L1455 have outflow energies that surpass their respective turbulent energies.  Their low virial parameters, however, suggest the outflows are not affecting the stability of the respective regions.  

Unlike other regions, IC348 and L1688 have low outflow energies relative to their turbulent kinetic and gravitational binding energies (i.e., $2E_\mathrm{out}/|E_\mathrm{grav}|\leq 0.06$).  Outflow feedback is unlikely to be the dominant driver of turbulence and has little effect on the dynamics of these regions.  With low turbulent kinetic energies as well (i.e., $2E_\mathrm{turb}/|E_\mathrm{grav}|\leq 0.3$), it is possible both regions are collapsing.

We note this analysis compares volumetric terms in the virial equation (neglecting surface terms).  Past work (e.g. \citealt{2006MNRAS.372..443B}) suggests clouds can be ram pressure confined, causing a region like Serpens Main to be bound even though it is super-virial.  We will address this in future using the full GBS dataset (Section~\ref{future}).


\subsection{Regional and Outflow Energetics}

\begin{figure}
\centering
\includegraphics[width=2.5in]{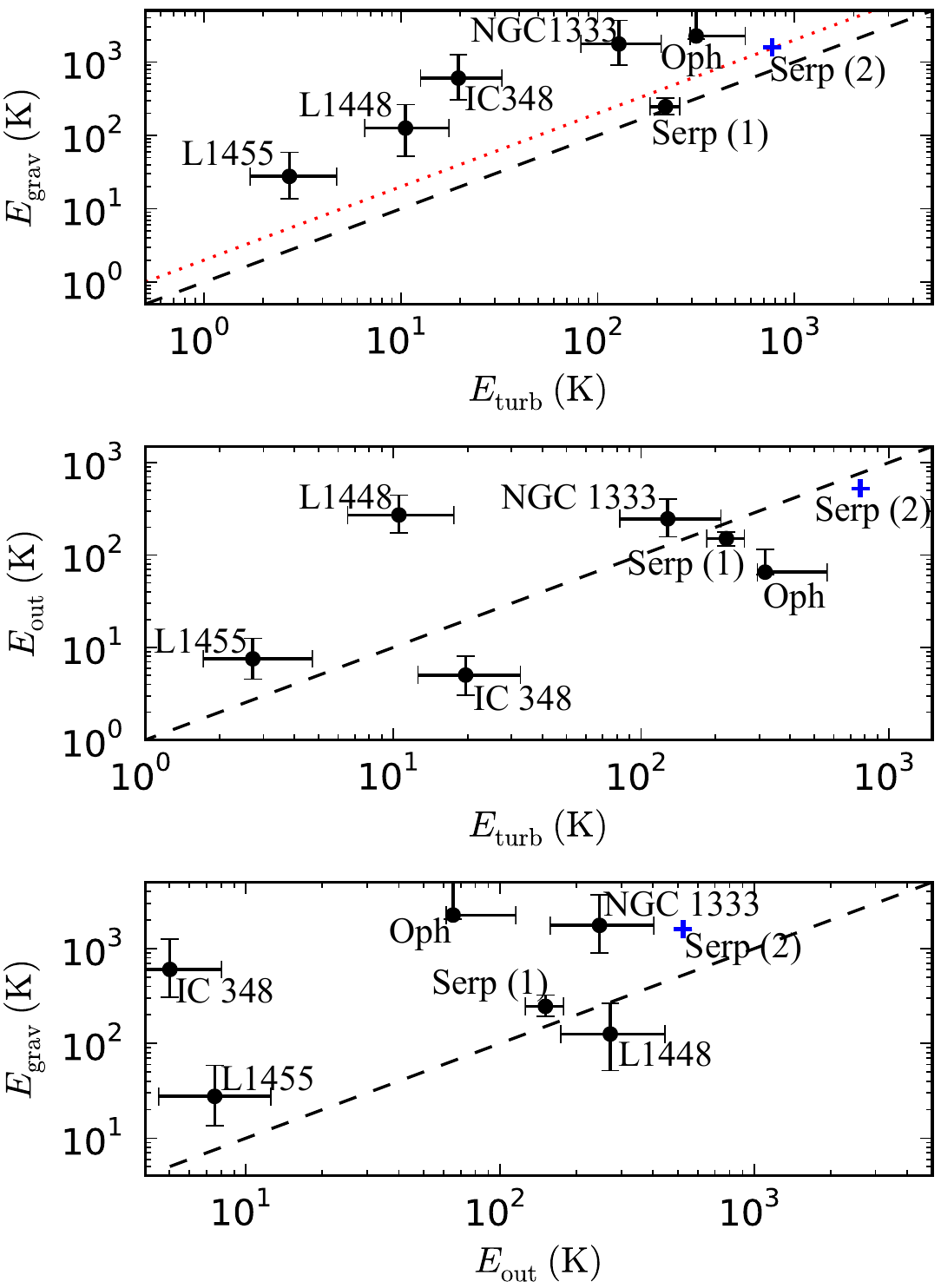}
\caption{Turbulent kinetic, gravitational binding and outflow energies.  {\emph{Top:}}  $E_\mathrm{turb}=E_\mathrm{grav}$ (dashed line) and $2E_\mathrm{turb}=E_\mathrm{grav}$ (dotted line).    {\emph{Centre:}} $E_\mathrm{turb}=E_\mathrm{out}$.  {\emph{Bottom:}} $E_\mathrm{out}=E_\mathrm{grav}$. Serp~(2) is denoted by the blue `$+$.' }
\label{fig:energy_comparison}
\end{figure}

Figure~\ref{fig:energy_comparison} shows a comparison of outflow, turbulent kinetic and gravitational energies.   We find a positive correlation between $E_\mathrm{turb}$ and $E_\mathrm{grav}$ with regions with higher turbulent energy also showing an increase in gravitational energy.  This is consistent with star-forming clouds being close to virial equilibrium (Figure~\ref{fig:mass_virial}).  Conversely, we do not find correlations between $E_\mathrm{out}$ and  $E_\mathrm{turb}$ or  $E_\mathrm{grav}$.  Even if outflows are generating turbulence, this lack of correlation indicates they are not the dominant sources determining turbulence levels.  However, not all the regions are the same, with some regions clearly having larger turbulent-to-outflow kinetic energies.  Regions with higher $E_\mathrm{turb}/E_\mathrm{out}$ may be more evolved since they will have fewer Class 0/I protostars to drive powerful outflows (\citealt{1996A&A...311..858B}; see Section~\ref{discussion}).  Additionally, since there is no relation between $E_\mathrm{out}$ and $E_\mathrm{grav}$, star formation (as measured by the outflow energy) does not appear to depend solely on large quantities of bound gas (see, however, \citealt{2014ApJ...787L..18S}).

\subsection{Injection and Dissipation Rates}

\begin{figure*}
\centering
\includegraphics[width=4.5in]{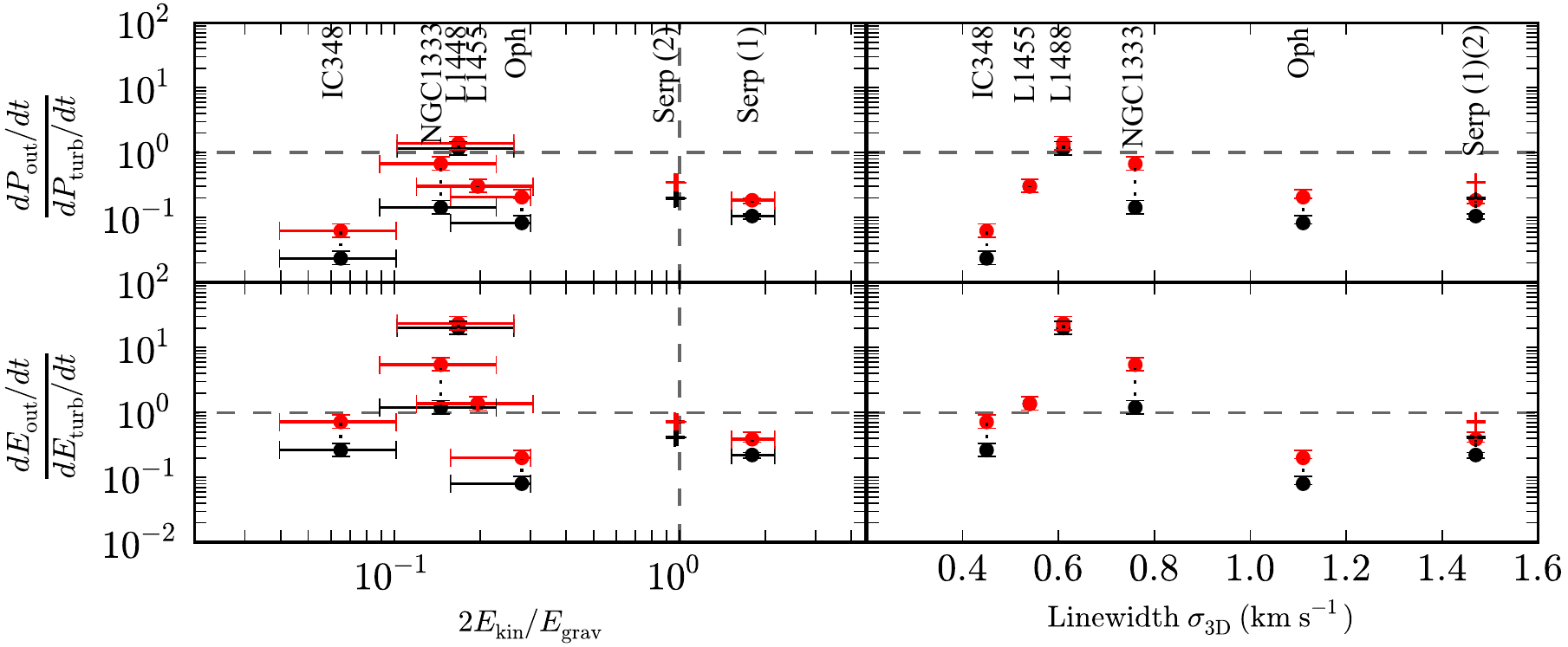}
\caption{Outflow injection and turbulence dissipation rates compared to the virial parameters (left) and velocity dispersions (right).  {\emph{Top:}}  Momentum injection/dissipation rates. {\emph{Bottom:}}  Energy injection/dissipation rates.  Serp (2) is denoted as `$+$'.}
\label{fig:rates}
\end{figure*}

Figure~\ref{fig:rates} shows the ratios of the outflow momentum and energy injection rates to the turbulence momentum and energy dissipation rates, compared to the virial parameter and the velocity dispersion, to understand how outflows affect the internal motions and stability of the ambient gas. We use a dissipation rate range assuming driving scales from that of an outflow (0.2~pc; average outflow length in Perseus, Serpens Main and Oph L1688 regions) to an effective regional radius.  Since L1455 has an effective radius of 0.15~pc, less than our assumed outflow length, we provide one estimate for its momentum and energy dissipation rates.  

Only L1448 has an outflow momentum injection rate greater than its turbulence dissipation rate.  NGC~1333, L1448 and L1455 have (average) ratios of momentum and energy injection and dissipation rates that are near or greater than unity, indicating outflows have enough energy to drive turbulence.  IC~348, L1688 and Serpens Main all have relatively low ratios of outflow momentum and energy injection rates.  There is no correlation between the rate ratios and the virial parameter or the velocity dispersion.


\section{Discussion and Conclusions}
\label{discussion}

How significant are outflows in driving turbulence in star-forming regions?  There is no doubt that outflows transport large amounts of energy and momentum.  The energy contained in outflows is greater than the total turbulent energy in half of the regions studied and at least 20~per~cent in all of the regions (Table~\ref{table1} and Figures~\ref{fig:mass_virial} and \ref{fig:energy_comparison}).  

If outflow activity increased cloud turbulence, then we would expect a correlation between the turbulent and outflow kinetic energies.  This does not seem to be the case (Figure~\ref{fig:energy_comparison}).  Similarly, if outflows had a significant effect in increasing turbulence, we might expect some of the regions to be close to the boundary of virial stability (preselection of active star-forming regions rules out gravitationally unbound examples).  However, the majority of the regions are strongly gravitationally bound, i.e. $2E_\mathrm{turb}/|E_\mathrm{grav}|\leq0.3$ (Figure~\ref{fig:mass_virial} and Table~\ref{table1}).  The one region that is borderline unstable is Serpens Main, which has a virial parameter close to unity and an outflow energy a factor of two less than its gravitational potential energy (as pointed out by G10).

We are left with the weaker possibility that outflows maintain turbulence levels by replenishing the energy and momentum that is dissipated by radiative shocks at small scales (e.g. \citealt{2010ApJ...722..145C}).  The strongest cases for outflow injection are Perseus regions NGC~1333, L1448 and L1455, where the average momentum injection-to-dissipation rates are close to unity (within a factor of $\sim3$), which is characteristic of the outflow-regulated cluster scenario \citep{2011ApJ...726...46N}.  Moreover, their energy injection rates exceed their turbulence dissipation rates, which could indicate that outflows are powerful enough to renew turbulence.  However, the argument for turbulence renewal by outflows weakens if some fraction of the outflow energy and momentum falls outside the cloud, particularly when the structure of the star-forming region is filamentary.  The {\emph{Herschel}} Gould Belt Survey estimated the average star-forming filament width to be $\sim0.1$~pc \citep{2014ASSP...36..259A}.  Comparing half of a filament width (0.05~pc) to our assumed 0.2~pc outflow length indicates that $\sim1/4$ of the momentum and energy is injected into the dense ambient gas where the turbulent energy is calculated.  This causes the lower bound of the momentum and turbulence injection-to-dissipation rates from Figure~\ref{fig:rates} to decrease by a factor of 4, suggesting outflows are not the main driver of turbulence.

There is evidence for a link between outflow contributions to turbulence and the evolutionary state of the star-forming region.  L1688, IC~348 and Serpens Main have lower ratios of energy and momentum injection to turbulence dissipation rates and lower ratios of $E_\mathrm{out}/E_\mathrm{turb}$.  These regions also have fewer Class~0/I YSOs (able to drive outflows) compared to the total YSO count (\citealt{2007ApJ...669..493W, 2008ApJ...683..822J}) at $<20$~per~cent, compared to NGC~1333 (35~per~cent) and L1455/L1448 (100~per~cent).

The relatively low outflow energies are unsurprising in L1688, which contains few Class~0 protostars (e.g. VLA1623; \citealt{2000prpl.conf...59A,Evans09}).   \citet{Evans09} noted the Ophiuchus cloud seems to have a declining star formation rate, indicated by a low number of Class~0 protostars and a higher YSO-to-cloud mass.  This decline in star formation will have reduced the outflow-driven turbulence in the region (low injection rates and higher $E_\mathrm{turb}/E_\mathrm{out}$).  Therefore, the turbulence preventing further collapse may result from winds generated by the Upper Sco OB association \citep{1989A&A...216...44D, 1990A&A...231..137D, 2012ApJ...754..104H}. 

IC~348 is a remnant of a larger cloud that formed the IC~348 cluster and the associated `Flying Ghost Nebula' \citep{1995A&A...300..276B}.  The stars that now produce this nebula are likely the sources of past outflows that created strong velocity gradients.  This region is estimated to have a declining star formation rate \citep{2007AJ....134..411M}.  The cores are primarily starless (three Class~0 protostars) and will likely go on to collapse and form protostars (suggested by C10).  Like L1688, the low outflow-to-turbulent energy in IC~348 is likely due to the lack of embedded protostars, where turbulence may be driven by winds from the IC~348 cluster.  

In conclusion, we find no evidence that outflows increase the turbulence levels in star-forming regions, though outflows may contribute significantly to maintaining levels of turbulence against dissipation in some cases.  Other mechanisms, such as supernovae or the process of cloud formation and subsequent mass growth (e.g. \citealt{2010A&A...520A..17K}), must be the main drivers of turbulence in most if not all of the star-forming regions.

\subsection{Comparisons to Past Work}
\label{future}
In the outflow-regulated cluster formation scenario, past work (e.g., \citealt{2014ApJ...783..115N}) suggested (1) the turbulence momentum dissipation rate must balance the outflow momentum injection rate and (2) the region must be close to virial equilibrium.  To test this, \citet{2014ApJ...783..115N} used line emission from CO isotopologues, where $^{12}$CO~$J=3\rightarrow2$ and $1\rightarrow0$ were used to trace outflow properties and $^{13}$CO~$J=1\rightarrow0$, C$^{18}$O and N$_2$H$^{+}$~$J=1\rightarrow0$ were used to trace ambient properties in regions that partially overlap with our analysis (B59, L1551, L1641N, Serpens Main, Serpens South, L1688, IC~348 and NGC~1333).  Their study finds virial parameters close to unity, except in Serpens South and L1688.  Outflow momentum injection rates were comparable to or larger than dissipation rates.  

Contrary to \citet{2014ApJ...783..115N}, our results suggest the majority of our regions have low virial parameters and low outflow momentum injection-to-dissipation rates for L1688, IC348 and Serpens Main.  Our dissimilar findings may be the result of using longer outflow timescales (i.e., Class~I age instead of dynamical timescales).  Additionally, our analysis uses C$^{18}$O~$J=3\rightarrow2$, which traces denser gas than $^{13}$CO~$J=1\rightarrow0$ and leads to lower cloud masses, radii and velocity dispersions.  This could result in regions having lower virial parameters and injection-to-dissipation ratios.  We note that $^{13}$CO~$J=1\rightarrow0$ from \citet{2014ApJ...783..115N} has the same line width as our C$^{18}$O data in L1688 even though $^{13}$CO traces a larger area and mass.  The lower $^{13}$CO line width is the result of averaging line widths from small sections in L1688 to obtain the global velocity dispersion (\citealt{1989ApJ...338..902L}).  Lastly, $^{12}$CO~$J=1\rightarrow0$ is better at detecting high-velocity outflowing material at lower densities, which could predict higher outflow energies than our study.  

In the future, we plan to address discrepancies in methods for calculating the outflow and ambient gas energetics.  We will also extend the study to other regions observed by the GBS, e.g. Orion A, Orion B and Serpens South.  




\section{Acknowledgements}

We would like to thank the referee for their time and helpful suggestions.  ED would also like to acknowledge funding from the Science and Technology Facilities Council of the UK.  

\bibliography{bib_energy}

\end{document}